# Safeguarding Scientific Integrity: Examining Manipulation in the Peer Review Process

Leslie D. McIntosh[1], Cynthia Hudson Vitale[1,2]
[1]Digital Science; [2]Association of Research Libraries

## Abstract
This case study analyzes the expertise, potential conflicts of interest, and objectivity of editors, authors, and peer reviewers involved in a 2022 special journal issue on fertility, pregnancy, and mental health. Data were collected on qualifications, organizational affiliations, and relationships among six papers' authors, three guest editors, and twelve peer reviewers. Two articles were found to have undisclosed conflicts of interest between authors, an editor, and multiple peer reviewers affiliated with anti-abortion advocacy and lobbying groups, indicating compromised objectivity.

This lack of transparency undermines the peer review process and enables biased research and disinformation proliferation. To increase integrity, we recommend multiple solutions: open peer review, expanded conflict of interest disclosure, increased stakeholder accountability, and retraction when ethical standards are violated. By illuminating noncompliance with ethical peer review guidelines, this study aims to raise awareness to help prevent the propagation of partisan science through respected scholarly channels.

## Introduction

Science is "the intellectual and practical activity encompassing the systematic study of the structure and behavior of the physical and natural world through observation and experiment" (Oxford Languages, 2022). Science, in theory, is inherently objective and based on a rigorous analysis of empirical evidence. Conversely, politics is a set of activities based on the subjective values of the majority governing group within a nation, state, or other area. Yet, neither science nor politics function in the abstract; both are human endeavors - influenced by human biases (Greenwald & Krieger, 2006)

The scientific method is one mechanism to check and balance personal biases and political influences on and within science (Gauch et al., 2003). The scientific method demands a skeptical mindset wherein the burden is placed upon the research to find sufficient evidence to prove their claim. Typically, that involves conducting experiments, collecting data, and presenting results of 'science' to provide evidence towards a consensus of truths for communication and discussion (Oxford Languages, n.d.)

Politics ostensibly operates in a parallel universe. Scientific and economic decisions and policies are balanced with the general population's needs and influenced by interested parties. In reality, science and politics have always been intimately connected (Howe, 2020), and neither works in practice as they do in theory. Science is political, and although politicians and lobbyists may not use the scientific method, they do use science. Politicians should make "evidence-based" decisions in the best interest of their constituents. This requires referring to empirical evidence, which is central to the scientific method. While science may be used politically, it is crucial to ensure that politics *does not interfere* with scientific methods or communications.

Scientific communication is informing, educating, and raising awareness of science-related topics (Burns et al., 2003). It is the premier mechanism for sharing information that may inform our society's policies, clinical care, and other critical areas. Scientific communications include established activities such as writing a scientific paper, presenting findings, participating in science policy and advocacy, and engaging in scholarly publishing through such endeavors as being a journal editor and participating in peer review (Hurd, 2000).

Despite these safeguards, established processes, and scholarly norms, research integrity, ethical lapses, and undisclosed conflicts of interest have been frequently documented. The primary question is to understand how mis/disinformation gets into science and how that information propagates outside of academia. To understand this, we can look at the gold of academia and scientific communications - publications - specifically the publication and editorial workflow.

## Communication through Publications: Editors, Reviewers, Authors, and the Publication Process

Peer review acts as one check of scientific worth and value, so let us start with evaluating the process and power of peer review. Introduced approximately fifty years ago, the peer review process, as we recognize it today, works when knowledgeable, independent experts comment on the quality of scientific research and uphold a set of ethical principles when evaluating manuscripts.

Ethical guidelines for peer reviewers have been established by the Committee on Publications Ethics (COPE). These guidelines articulate three areas of importance for editors and peer reviewers to consider before accepting an assignment: professional responsibility, competing interests, and timeliness. Digging into professional responsibility and competing interests a bit more, one finds the following ethical considerations (Committee on Publication Ethics, n.d.):

1. Expertise: Editors and peer reviewers should have the relevant expertise and knowledge to properly evaluate the work.
2. Conflict of Interest: Editors, peer reviewers, and authors should declare competing interests, which "may be personal, financial, intellectual, professional, political, or religious in nature."
3. Objectivity: Editors and peer reviewers should provide impartial and fair evaluations of the work under review.

Scientific journal editors select peer reviewers. With the appointment of guest editors, the power of peer review selection and oversight of publications moves from the journal editor to the appointed guest. It is imperative that the guest editors adhere to ethical and publisher guidelines. If they breach these guidelines, it may result in biased research communicated through a trusted channel. Breaching guidelines for political or professional gains is a highly concerning situation as journals and their board of editor's reputations can all be called into question.

In theory, all reviewers should be independent of one another, but in practice, this is rarely the case. 'Peers,' by definition, means there will be some overlap among scientific publishing networks and the expertise of individuals. The editors, peer reviewers, or authors are often part of the same scientific society.

Because the peer review process and workflow can vary by publisher and journal (Brezis & Birukou, 2020) and a common peer review workflow has not been codified, the peer review processes do not fully guarantee research integrity. Additionally, the difference between gaming and manipulating peer review may not be clear. The first is a grey area of knowing how the system works and optimizing the process for professional gains (e.g., promoting your work to gain citations). The latter refers to knowing how the system works and stepping over community boundaries of acceptable practices. Guidelines inform and lead the community in what is considered ethical and acceptable practices.

These ethical practices, as defined by COPE and others, are meant to ensure the reliability and validity of the peer-review process and to ultimately promote the advancement of knowledge in a particular field. The International Committee of Medical Journal Editors (ICMJE) further articulates these ethical practices by indicating that peer reviewers must "disclose to editors any relationships or activities that could bias their opinions of the manuscript, and should recuse themselves from reviewing specific manuscripts if the potential for bias exists" (ICMJE, n.d.). Many publishers and journals adhere to the COPE and ICMJE guidelines for peer review and, in some cases, go beyond them (Frontiers, n.d.; Jefferson et al., 2002).

These organizations (COPE and ICMJE) guide publishers, authors, and institutions but cannot fully protect scientific integrity. Through shared beliefs of searching for 'truth,' there are expectations from all stakeholders - from researchers to government - to abide by honest practices of accepted norms (Resnik & Elliott, 2023). When these norms are not upheld, research integrity issues may arise, including conflicts of interest.

## Study Objective

The objective of this study is to present a case study on how the peer review process may be manipulated by individuals with undisclosed connections to politically divisive organizations. To do this, we leverage the COPE ethical guidelines for peer reviewers to document relationships among peer reviewers - specifically focusing on editor and reviewer expertise, conflict of interest for authors, peer reviewers, and editors, and objectivity. This case study analyzes these relationships in one special issue of a journal.

# Materials and Methods

## Data Collection

Articles for analysis were identified using these criteria: 1) All articles are included in the same special issue with the same guest editors; 2) Published in 2022; 3) Available open access; and 4) Peer-reviewed with peer-reviewer names publicly available.

Article authors, editors, and peer reviewers' affiliation information were gathered from the Frontiers special topic web page, web searches, and the Dimensions database (Digital Science, 2018). Each person's role in the special issue was collected from the Frontiers Research Topic website in February 2023.

## Expertise

To evaluate expertise, we collected the number of prior publications through Dimensions, the last year of publication, and the number of peer reviews using ORCiD (ORCID, n.d.) and Loop (Frontiers, n.d.). To quantify expertise as guest editors and peer reviewers, we collected the total number of publications, average publications over four years (2019-2022) with fewer than 25 co-authors (thus eliminating mega-author publications), and the last year of a publication (up to 2022). This gave us a proxy for understanding their expertise and qualifications as peer reviewers.

## Conflicts of Interest

For assessing conflicts of interest, we are guided by the ICMJE definition wherein "all participants in the peer-review and publication process – not only authors but also peer reviewers, editors, and editorial board members of journals – must consider and disclose their relationships and activities when fulfilling their roles in the process of article review and publication." (ICMJE, n.d.).

Frontiers - the publisher of this special issue - has specific instructions on conflict of interest: "A conflict of interest can be anything potentially interfering with, or that could be perceived as interfering with, full and objective peer review, decision-making or publication of articles submitted to Frontiers. Personal, financial, and professional affiliations or relationships can be perceived as conflicts of interest." (Frontiers, n.d.).

To evaluate conflicts of interest (COI), we collected institutional and organizational affiliations by reviewing the author, peer reviewer, and editor's public presence (e.g., CV, organizational profile, LinkedIn) through web searches. Once authorship was confirmed, we mined organizational websites for author and reviewer affiliation and citations for the role of affiliation within the organization. Specifically for the COI of reviewers and authors, we cannot say if a person will be unbiased - we know any human in any position has biases. However, we can assess if there has been an attempt to curtail biases. One way is to look at the disclosure of conflicts of interest.

## Objectivity

Objectivity is defined by Resnik and Elliott (2023) as "minimizing or controlling experimental, theoretical, and other biases, including conflicts of interest."

For the purposes of this research, this variable is calculated by combining the expertise and conflict of interest data. Additional details on how this calculation was undertaken are detailed in the analysis section. No data collection of what was stated in the peer reviews of the articles was obtained and could not be used to assess the objectivity of their reviews.

## Analysis

A mixed-methods approach was applied for all data collected to analyze expertise data and the conflict of interest.

## Conflict of Interest

Once an organizational affiliation was identified, we assessed the strength of affiliation based on the following criteria. They have strong political affiliations if: i) they are a member, leader, or founder of one of the organizations; or ii) they are considered an advocate or activist if they write papers for political organizations affiliated with the topic or act as expert witnesses aligning with political beliefs.

## Objectivity

To assess objectivity, the authors conducted an exploratory network analysis using the Hevey model for network analysis that creates a visual representation of relationships (edges) between variables (nodes) (Hevey, 2018). This type of analysis is appropriate as it allows the visualization and analysis of a system of beliefs, emotional states, behaviors, and symptoms that influence each other over time.

More specifically, we applied a bipartite network analysis (Asratian et al., 1998) that visualizes the roles (editors, peer reviewers, authors) as the first set of nodes and papers as the second set of nodes. The directed edges link the people to the papers they edited, reviewed, or authored. We then color-coded the nodes with conflicts of interest to understand the distribution of COIs across the papers on the Frontiers research topic.

## Results

The special research issue (Sammut et al., n.d.) analyzed contained one overview from the guest editors and six articles subject to peer review (Table 1). Two other articles appeared on this special topic after that date. However, neither had the same guest editors and were excluded from this analysis. Note that we have presented the results in a deidentified manner so readers can focus on viewing the potential conflicts of interest.

The research topic overview included the three guest editors and is not considered an 'article' as it does not have a DOI, and thus it is not indexed in our database, Dimensions. Of the six articles, one is a 'brief research report,' two are 'reviews,' and three are 'original research.' One review (105) had four peer reviewers. The other articles each had two peer reviews. Two papers had an editor as one of the reviewers (101 and 103).

**Table 1: Articles in the *Frontiers* special research issue through June 2022**

| Paper ID | 100 | 101 | 102 | 103 | 104 | 105 | 106 |
|---|---|---|---|---|---|---|---|
| Paper Type | Research Topic | Original Research | Original Research | Review | Original Research | Review | Brief Research Report |

## Results: Expertise

Throughout the special issue, there were three guest editors and twelve other peer reviewers (Table 2). All editors and reviewers have appropriate levels of education and practice in an area related to the paper (data not shown). They have either doctorates in a related field or are medical practitioners. While Guest Editors 1 and 2 work in the area of the special topic, Guest Editor 3 is a neuroscientist with no work history in fertility, pregnancy, and mental health.

Three peer reviewers, while having longer careers, have few publications: reviewers 7, 10, and 11 had no publications in the last four years. The last year of a publication is striking for reviewers 7 and 10 as they have not been primary authors on a paper since 2015 and 2007, respectively. Reviewer 11 stands out with no publication history and is not an early career researcher. Reviewers 5 and 6 have remarkable publication records on the opposite spectrum - as they are both prolific authors.

We also attempted to evaluate expertise by how many peer reviews each had conducted in their recent professional career, using Loop (Frontiers researcher platform) and ORCiD. However, as those can both be incomplete and were entirely missing (i.e., no ORCiD, hence no peer reviews), we excluded the data from these sources in the table.

**Table 2: Publication history over four years (2019-2022) by guest editors and reviewers for the research topic.**

| ID | Role in Special Issue | Publications | Average Publications over four years | Year of Last Publication |
|---|---|---|---|---|
| 1 | **Guest Editor** | 22 | 1.3 | 2022 |
| 2 | **Guest Editor** | 43 | 1.5 | 2021 |
| 3 | **Guest Editor** | 11 | 0.5 | 2021 |
| 4 | **Reviewer** | 16 | 2.0 | 2022 |
| 5 | **Reviewer** | 335 | 17.0 | 2022 |
| 6 | **Reviewer** | 150 | 31.3 | 2022 |
| 7 | **Reviewer** | 24 | 0.0 | 2015 |
| 8 | **Reviewer** | 2 | 0.5 | 2021 |
| 9 | **Reviewer** | 22 | 3.8 | 2022 |
| 10 | **Reviewer** | 5 | 0.0 | 2007 |
| 11 | **Reviewer** | 0 | 0.0 | – |
| 12 | **Reviewer** | 4 | 1.0 | 2022 |
| 13 | **Reviewer** | 532 | 18.8 | 2022 |
| 14 | **Reviewer** | 43 | 7.3 | 2022 |
| 15 | **Reviewer** | 65 | 11.0 | 2022 |

## Results: Conflict of Interest

We identified multiple organizations associated with one or more of the editors, reviewers, or authors that are political and should be identified as a potential conflict of interest.

All affiliations appeared on public profiles, publications, or CVs: Charlotte Lozier Institute (CLI), American Association of Pro-Life Obstetricians and Gynecologists (AAPLOG), Breast Cancer Prevention Institute (BCPI), World Expert Consortium for Abortion Research and Educations (WECARE), St Louis Guild of the Catholic Medical Association. More information on why these are considered anti-abortion organizations can be found in the Appendix.

Our analysis found two of the three guest editors, five of the twelve (42%) peer reviewers, and two of nineteen authors (11%) are active in anti-abortion advocacy and lobbying organizations that are conflicts of interest:
1. Charlotte Lozier Institute (CLI)
2. American Association of Pro-Life Obstetricians and Gynecologists (AAPLOG)
3. Breast Cancer Prevention Institute (BCPI)
4. World Expert Consortium for Abortion Research and Education (WECARE)
5. St Louis Guild of the Catholic Medical Association

Guest Editor 3 has affiliations with Catholic organizations. However, there are no publications or organizations to understand his potential conflicts of interest on the topic. It should be noted that while Catholic organizations or individuals of the Catholic faith may have personal biases around abortion, they may not be actively lobbying or advocating for anti-abortion policies. Hence, this is not considered a conflict of interest.

There is another paper with all authors affiliated with Life Perspectives, San Diego, which is Christian-based. While CLI has cited these two authors twice, this organization does not appear to have connections with CLI or AAPLOG. Moreover, the peer-reviewers for this paper appear to not be associated with the authors or the author's organization (data not shown).

Of fourteen reviewers, eight are strongly connected with CLI, AAPLOG, Breast Cancer Prevention Institute, WECARE, or other anti-abortion organizations. The two authors with political activity in the area are single authors in their papers. Eight of the nine individuals with conflicts of interest did not declare them in the article (Table 3).

**Table 3: Stated and unstated conflicts of interest in the research topic *Fertility, Pregnancy, and Mental Health - a Behavioral and Biomedical Perspective***

| ID | Role | Charlotte Lozier Institute | American Assoc. of Pro-Life OBGYN | Breast Cancer Prevention Institute | World Expert Consortium for Abortion Research and Education | Expert Witness Supporting Anti-abortion Cases | Other Anti-Abortion Affiliation and Activities |
|---|---|---|---|---|---|---|---|
| 1 | Guest Editor | ■ undisclosed | ▨ cited | | ■ undisclosed | | |
| 2 | Guest Editor | | | | | | ■ undisclosed |
| 3 | Guest Editor | | | | | | |
| 7 | Reviewer | ▨ cited | ▨ cited | | ■ undisclosed | ■ undisclosed | |
| 10 | Reviewer | ■ undisclosed | | | | | |
| 11 | Reviewer | ■ undisclosed | ▨ stated | | | | |
| 12 | Reviewer | ▨ stated | | | | | |
| 13 | Reviewer | ▨ cited | ▨ cited | | | ■ undisclosed | ■ undisclosed |
| 16 | Author | ▨ cited | ■ undisclosed | ■ undisclosed | ■ undisclosed | | |
| 17 | Author | ▨ cited | ▨ cited | | ■ undisclosed | ■ undisclosed | |

Undisclosed affiliation in conflict of interest statement

Stated affiliation but not indicated in the conflict of interest statement

The organization cited* a person one or more times.

*Cited by CLI means the author who wrote or was cited in blog posts or other writings published by the Charlotte Lozier Institute. Note that being cited by CLI does not indicate an endorsement from the person being cited.*

## Results: Objectivity

Can we make a judgment based on objectivity with expertise, role (power), and COI, even without seeing the peer reviews? For this, we looked at the bipartite network.

The results of the exploratory bipartite network analysis show the six articles with connections to the editors (shield), authors (circle), and peer reviewers (triangles) (Figure 1). Guest Editor 1 edited five papers and reviewed one (101); Guest Editor 2 was neither editor nor reviewer on any papers; and Guest Editor 3 was the editor on two papers and reviewer for one (103).

We see several potential influences in the peer review process that conflict with the intended independent nature of the peer review process. Specifically, we found significant overlap in affiliation and authorship between the author, editor, and peer reviewers of two of papers 103 and 105 (Figure 2).

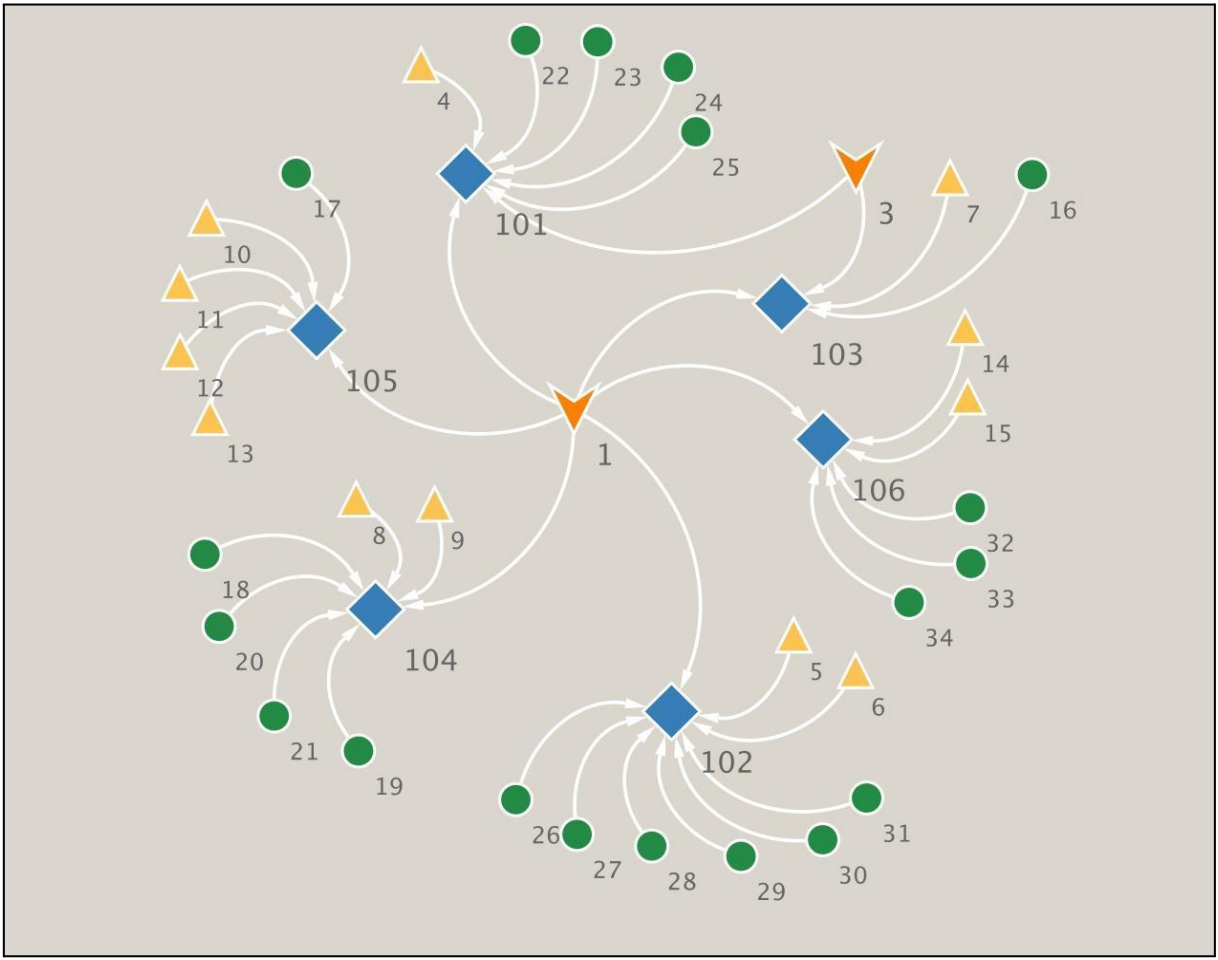

**Figure 1: Network showing connections between the editors' and reviewers' links to the special issue articles.**

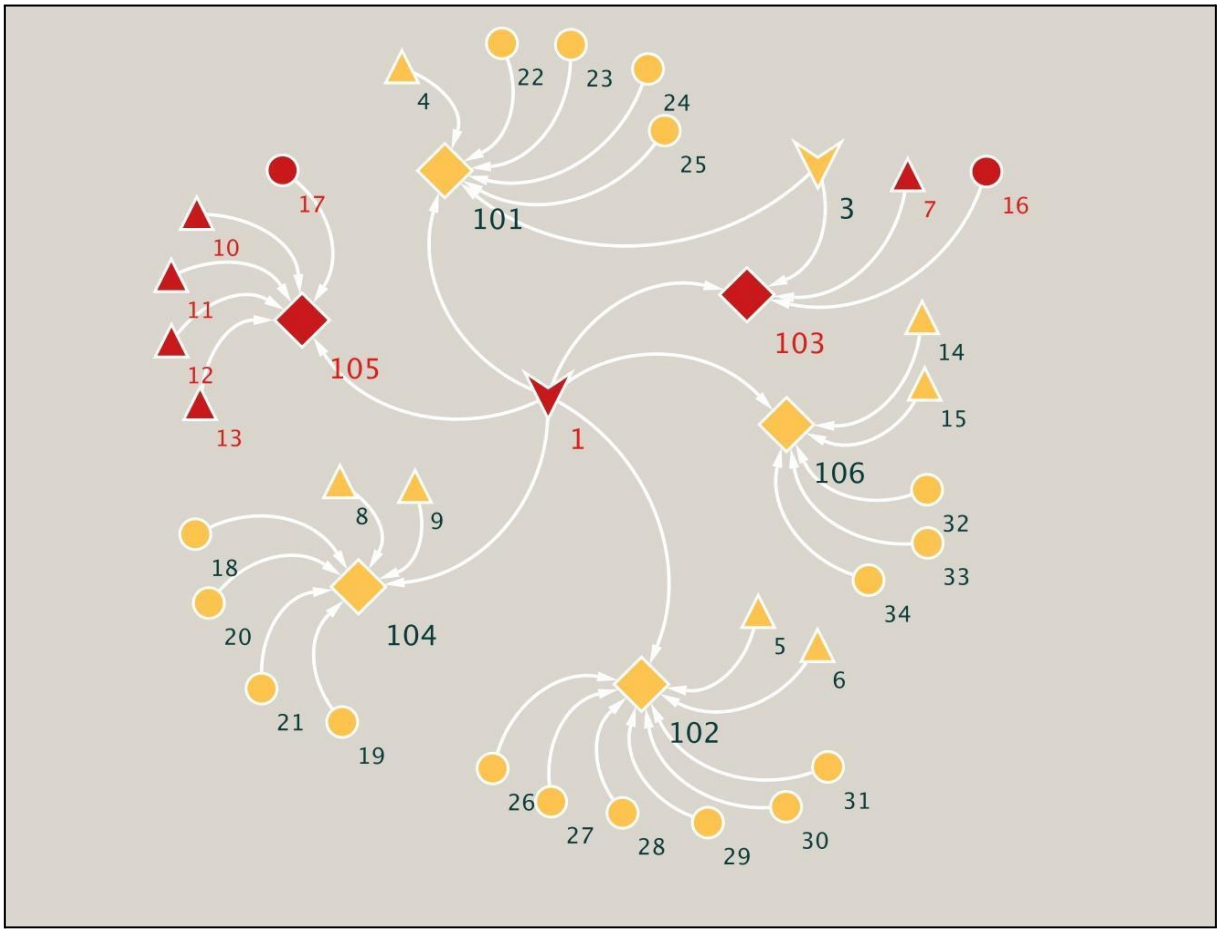

Figure 2: Network with the same connections in Figure 1, with the conflicts of interest color-coded. Those individuals with conflicts of interest are shown in red, while those without conflicts of interest are shown in yellow.

Network Legend

| Paper | Guest Editor | Reviewer | Author | Conflict of Interest |
|---|---|---|---|---|
| ◇ | ∀ | △ | ○ | 🟥 |

## Discussion

Our objective has been to present a case study on how the peer review process may be manipulated by individuals with undisclosed connections to politically divisive organizations. Conflicts of interest in this special issue should have prevented publication of two of the articles. Yet, this study was made possible because of the publishers's transparency of the guest editors and peer reviewer - magnifying the need for multiple checks of scientific integrity.

It is accepted practice for editors to select reviewers with expertise from different topics or methods knowledge (e.g., network analysis and genomics) and to favor those who have previously published. It is known that finding qualified and available peer reviewers is challenging, and an editor will act as a peer reviewer on occasion.

The results indicate that an ethical lapse may have occurred in two of the articles published in this special issue. We inferred that the guest editors and peer reviewers agreed to abide by the publisher and COPE guidelines when agreeing to be in their roles. Yet, looking at the ethical guidelines for peer reviewers as defined by COPE and ICMJE, one finds questions of objectivity and expertise and several instances of undisclosed conflicts of interest. There appear to be significant operations of aligning peer reviewers by a guest editor to support reviews of papers that align politically with their views and are not focused on diverse peer reviews. Hence, there appears to be significant bias in this special issue's publication.

One of the suspect papers identified has since been retracted due to "undisclosed competing interests…, which undermined the objective editorial assessment of the article during the peer review process" (Coleman, 2022). The retraction took months and involved legal proceedings (Marcus, 2022). The other paper has yet to be retracted, although it includes undisclosed and potentially questionable competing interests.

While we have highlighted two known conflicts of interest issues in this special issue, our analysis found no issues with the other papers in the collection. Four articles appear to have reviewers with the requisite expertise, no undisclosed conflicts of interest, and adequate objectivity with the authors. The negative attention and retraction of an article within this special issue may impact the authors of the other papers, but this is difficult to measure.

Other limitations to this study also exist: i) Collecting peer review information is difficult and would have added to this study; ii) We may have missed associations as there is no registry of organizations affiliations. An absence of finding associations does not indicate the absence of affiliation; and iii) This case study is on one research topic. The sample size of six papers is naturally small, and there are no comparisons with other special issues. Yet, other research has discussed many cases of manipulating special issues.

## Recommendations

All scholarly communications stakeholders play a role in increasing the quality and integrity of the research process. Unfortunately, it is unclear who should be responsible for the consequences of manipulating the system. The guest editors, who select the peer reviewers, have a clear responsibility. Yet, the guest editors practice at a journal and work at institutions, meaning there can be other checks and consequences within the scholarly publication workflow. The following recommendations are proposed as a series of mechanisms to further reduce confirmation bias and increase research integrity within two stakeholders, publishers and institutions.

## Open Peer Review

One measure to illuminate future biases of this type is adopting a more open peer review process to increase research transparency and the integrity of published research. In this case study, the Frontiers journal transparently posted the peer reviewers and editors. This

McIntosh and Hudson Vitale  11

transparency allowed us to review the potential conflicts of interest without requesting information from the publisher.

Open peer review has varying implementation models, including sharing peer reviewer names, as is the case here. But other models go further where "aspects of the peer review process are made publicly available before or after publication" (Besançon, n.d.). In practice, open peer review may also involve publishing the review letters from peer reviewers and editors either in their original form or as aggregated recommendations.

Because the guest editor and peer reviewer names were open, this helped identify anomalies in the guest editorial practices and peer review process. This transparency of authors and editors is critical for allowing researchers and the general public to conduct analyses and make informed decisions about the resulting research. However, we must go further.

In addition to reducing the confirmation bias, open peer review may support increased consistency among peer reviews, allow incentives for the peer review process to be developed, and make the "black box" of the traditional peer review workflow more transparent. To maximize transparency, publication metadata should be automatically transferred into open, authenticated ID systems, such as ORCiD.

## Conflict of Interest Database

We urge the scientific community to consider increasing the transparency of affiliations with lobbying groups, non-profits, and other organizations. Much like a conflict of interest statement, an affiliation disclosure for board positions and memberships would provide the public and policymakers with additional information to assess any potential biases of the research outcomes.

A handful of publishers are more specific in their editorial policies, requiring authors to declare financial and non-financial interests. While financial interests, such as employment or personal financial interests, that may gain or lose financially through the publication are common areas of disclosure, non-financial are less common. Non-financial interests may include non-paid membership in a government or non-governmental organization, an advisory position in a commercial organization, and membership in an advocacy or lobbying organization. The publisher of this special issue does require all editors and reviewers to disclose personal conflicts of interest (such as family), collaborations (such as recent, previous authorship with the author of the article to be published), financial conflicts of interest, and affiliations (such as working at the same institution as the author of the article to be published). As this research has shown, publishers would benefit from adopting more rigorous disclosure policies and verifying their accuracy.

## Take Responsibility

Publishers and institutions could require training for editors, peer reviewers, and authors on ethical practices for peer review and mandate that every author on a manuscript have a persistent author identifier (e.g., ORCiD). Institutions should hold responsibility for the researchers they employ. They can also modify the promotion and tenure review criteria to increase incentives for ethical and open peer review practices.

Manuscripts should not be scientifically edited or reviewed by peer reviewers who cannot meet ethical guidelines. Moreover, if the ethical guidelines are not met, the papers should have a notice placed on them that the ethics should be questioned - in this case due to a non-disclosure of COIs. If a submitted paper would have been denied review due to COIs but was none-the-less published, the articles should be retracted.

Acting honestly, the author should withdraw any papers with unstated COIs, and those papers should not be used in making policies.

## Conclusion

Any time established publishers from experienced researchers distribute questionable research, those articles and the results can be evidence for federal, state, and local policies and laws.

We intentionally did not assess the quality of the research and scientific processes within these articles because we questioned the quality of the editorial process and workflow that allowed the articles to be published and had grave concerns about the downstream effects. This type of bias is particularly problematic given the potential impact on policies and case law. The toxins embedded in the two biased articles create secondary pollution from outside the scientific ecosystem. All stakeholders must be held accountable, and these informational pollutants must be curtailed. As disinformation proliferates, mitigating its spread into the scholarly literature is critical to safeguard society from its broader use.

## Acknowledgments

The authors wish to thank Daniel Hook, Simon Linacre, and David Ellis for the reviews and comments on the article. Thank you to Simon Porter for thinking through and assisting with some of the network analysis and the persnickety Cytoscape software.

## Conflict of Interest

*Susan B Anthony List Inc Education Fund*. (n.d.). GuideStar. Retrieved August 7, 2023, from https://www.guidestar.org/profile/26-4788700

Wikipedia. (2023). *Science communication*. https://en.wikipedia.org/wiki/Science_communication

*World Expert Consortium for Abortion Research and Education*. (n.d.). Wecare. Retrieved August 7, 2023, from https://www.wecareexperts.org/affiliates/

# APPENDIX

Organizations identified as affiliated with an editor, peer reviewer, or author that may be interpreted as a conflict of interest.

Conflict of Interest: "The potential for conflict of interest and bias exists when professional judgment concerning a primary interest (such as patients' welfare or the validity of research) may be influenced by a secondary interest (such as financial gain). Perceptions of conflict of interest are as important as actual conflicts of interest." (ICMJE, n.d.)

1. **Susan B. Anthony List Education Fund, d/b/a Charlotte Lozier Institute (CLI),** is the education and research arm of Susan B. Anthony List. CLI is a non-profit [501(c)(3) public charity.](#) Their primary program is "Education and Research - CLI continued activities as the preeminent organization for science-based pro-life information and research." The Susan B. Anthony List Education Fund describes their primary activity as "to end abortion via activist training and passage of legislation ."(*Susan B Anthony List Inc Education Fund*, n.d.)

2. **American Association of Pro-Life Obstetricians and Gynecologists (AAPLOG)** functions as a professional association offering "resources, medical education opportunities, and support to our various constituencies. We inform professionals in the medical and legal arenas via effective medical life-affirming arguments. Key partners: Alliance Defending Freedom, Americans United for Life, the Lozier Institute, the Becket Fund and other defenders of liberty and conscience. " (*Am. Assoc. of Pro-Life Obstetricians and Gynecologists*, n.d.)

3. **Breast Cancer Prevention Institute (BCPI)**
   The Breast Cancer Prevention Institute is a non-profit 501(c)(3) corporation which educates healthcare professionals and the general public through research publications, lectures, and the Internet on ways to reduce breast cancer incidence. "BCPI helped an over 6,000 member strong physician association, the American Association of Pro-Life Obstetricians and Gynecologists (AAPLOG), to document an amicus curiae brief that abortion negatively impacts women's health and future children." (*BCPI Resources*, n.d.)

4. **World Expert Consortium for Abortion Research and Education (WECARE)**
   They are not discoverable as a non-profit through Guidestar. "WECARE brings together credentialed scientists with a research program on the physical, psychological, and/or relational effects of abortion on women and those closest to them to engage in international research collaboration, scientific information dissemination, professional education, and consultation." (*World Expert Consortium for Abortion Research and Education*, n.d.)